\documentclass[prl,twocolumn,showpacs]{revtex4}

\usepackage{graphicx}
\usepackage{amsmath}
\usepackage{amssymb}
\usepackage{epsf,latexsym}
\usepackage{times}

\newcommand{\be}{\begin{equation}}
\newcommand{\ee}{\end{equation}}
\newcommand{\ba}{\begin{eqnarray}}
\newcommand{\ea}{\end{eqnarray}}

\newcommand{\putfig}[2]{$$\leavevmode\hbox{\epsfxsize=#2 cm
   \epsffile{#1.eps}}$$}

\begin{document}


\title{Universal scaling of Coulomb drag in graphene layers}

\author{Irene D'Amico$^{1,2}$}
\email{damico@isiosf.isi.it}

\author{S.G.~Sharapov$^{1,2}$}
\email{sharapov@bitp.kiev.ua}


\affiliation{$1$ Institute for Scientific Interchange, via Settimio Severo 65,
I-10133 Torino, Italy\\
        $^2$Istituto Nazionale per la Fisica della Materia
        (INFM), Italy}

\date{\today }

\begin{abstract}
We study the Coulomb drag transresistivity between graphene layers
employing the finite temperature density response function. We
analyze the Coulomb coupling between the two layers and show that
a universal scaling behavior, independent of the interlayer
distance  can be deduced. We argue that this universal behavior
may be experimentally observable in a system of two carbon
nanotubes of large radius.
\end{abstract}







\maketitle Coulomb interactions are responsible for a rich variety
of phenomena in low-dimensional systems. In particular, there are
many indications that in semi-metals, such as graphite, the
Coulomb forces remain long ranged due to the lack of conventional
screening. This property of graphite has intensified efforts from
both experimental \cite{Kopelevich:2002} and theoretical
\cite{Gonzalez:1999:PRB,Gonzalez:2001:PRB,Khveshchenko-Gorbar}
sides to understand its electronic properties and
interaction-driven transitions. However, a direct linear-response
transport measurements of the effect of electron-electron
interactions in single  isolated sample are usually difficult
because for a perfectly pure, translationally invariant system,
the Coulomb interaction cannot give rise to resistance. It has
been demonstrated \cite{Pogrebinskii-Price} that when two
independent electron gases with separate electrical contacts are
placed in close proximity and current is driven through one, the
interlayer electron-electron interaction creates a frictional
force which drags a current through the other. This so-called
Coulomb drag effect \cite{Rojo:1999:JPCM} is unique in that it
provides an opportunity to measure directly electron-electron
interactions.

The purpose of the present paper is to demonstrate that the
semi-metal band structure of graphene is responsible for an
unusual behavior of the Coulomb drag transresistivity. In addition
our calculations show a universal scaling behavior of the Coulomb
drag transresistivity, which
may be experimentally observable.

It is  convenient to characterize the Coulomb drag effect
in terms of the transresistivity $\rho_{21}$ because in a drag-rate
measurements $\rho_{21}$ is directly related to the rate of momentum
transfer from particles in layer 1 to layer 2. However,
when Kubo formalism is employed \cite{Kamenev-Flensberg}
one arrives at the transconductivity $\sigma_{21}$.
These two quantities are defined as $\rho_{21} = {E_2}/{j_1}$ with $j_2 =0$ and
$\sigma_{21} = {j_2}/{E_1}$ with $E_2 =0$,
where $E_i$ and $j_i$ are, respectively, the electric field
and the current density in  layer $i$. Since the transconductivity
is caused by a screened interaction between spatially separated layers,
$\sigma_{21} \ll \sigma_{11}$ and $\rho_{21}$ is related to $\sigma_{21}$
via $\rho_{21} \approx - \sigma_{21}/(\sigma_{11} \sigma_{22})$.

The expression for the dc Coulomb drag transresistivity in two dimensional
systems, in the RPA approximation, is given by
\cite{Rojo:1999:JPCM,Irene:2003:PRB}
\begin{equation}\label{transresistivity.def}
\begin{split}
& \rho_{21}(T) =
 -\frac{\hbar^2}{8\pi^2 e^2 k_B T n_1 n_2} \int_{0}^{\infty} d K
K^3  \\
& \times \int_{0}^{\infty} d \omega \left|V_{12}(\omega,K)\right|^2\frac{\mbox{Im}
\chi_{1}(\omega,K) \mbox{Im}
\chi_{2}(\omega,K)}{\sinh^2
\frac{\hbar \omega}{2 k_B T}}.
\end{split}
\end{equation}
Our purpose is   to describe Coulomb drag between two {\it
graphene} layers, i.e. between two planar sheets of carbon atoms,
so that Eq.~(\ref{transresistivity.def}) is purely two
dimensional \cite{Irene:2003:PRB}.

In Eq.~(\ref{transresistivity.def}) $V_{12}(\omega,K) = \exp(-d K)
v(K)/\epsilon(\omega,K)$ is the screened interlayer interaction,
$v(K) = 2 \pi e^2/(K \varepsilon)$ is the Fourier transform of the
Coulomb interaction with the appropriate low-frequency dielectric
constant $\varepsilon$, $\chi_i(\omega,K)$ is the noninteracting
density-density response function (the ``Lindhard function''),
$n_i$ is the total carrier density, $d$ is the interlayer
distance, $e$ is the electron charge and $i=1,2$ is the layer
index. The RPA dielectric function $\epsilon(\omega,K)$ can be
written as \cite{Rojo:1999:JPCM} \be \label{epsilon}
\begin{split}
\epsilon(\omega,K)  =  [1-v(K) \chi_1(\Omega,K)]
[1-v(K) \chi_2(\Omega,K)]\\- \exp(-2 d K))v(K) \chi_1(\Omega,K)v(K) \chi_2(\Omega,K).
\end{split}
\ee
We stress that
since in contrast to the total
current $j(\mathbf{K}=0)=
j_{1}(\mathbf{K}=0)+j_{2}(\mathbf{K}=0)$, the current in a given
layer
 $j_{i}(\mathbf{K}=0)$ is
not conserved, the transresistivity
Eq.~(\ref{transresistivity.def}) is given by fluctuation diagrams
(see Refs.~\cite{Kamenev-Flensberg}), which are similar but not
identical to the Aslamazov-Larkin diagram known from
superconductivity \cite{Larkin.book}, and not by a simple bubble
as for the conventional electrical conductivity. Thus the
transresistivity appears to be sensitive to all kinds of virtual
excitations that exist in the system whenever $\mbox{Im}
\chi(\omega, K)$ is nonzero.

Let us first of all derive the Lindhard function for graphene sheets {\it including}
finite temperature effects.
For a single graphene layer the Lindhard function may be written as
\begin{equation}\label{LF.res}
\chi (i\Omega, \mathbf{K}) =
\chi_+ (i\Omega, \mathbf{K}) +
\chi_- (i\Omega, \mathbf{K})
\end{equation}
with
\begin{equation}\label{LF.res+}
\begin{split}
& \qquad \chi_+ (i\Omega, \mathbf{K}) =
- 2 \int \frac{d^2 k}{(2 \pi)^2} \times \\
& \left\{
\frac{A_{+}}{E_{+} - E_{-} + i \hbar\Omega}
[n_{F}(\mu + E_-) - n_F(\mu + E_+)] \right. \\
& \left. + \frac{A_{+}}{E_{+} - E_{-} - i \hbar\Omega}
[n_{F}(\mu- E_+) - n_F(\mu - E_-)]  \right\},
\end{split}
\end{equation}
and
\begin{equation}\label{LF.res-}
\begin{split}
& \qquad \chi_- (i\Omega, \mathbf{K}) =
- 2 \int \frac{d^2 k}{(2 \pi)^2} \times \\
& \left\{
\frac{ A_{-}}{E_{+} + E_{-} + i \hbar\Omega}  [
n_{F}(\mu- E_-) - n_F(\mu + E_+)]\right.  \\
& \left. + \frac{ A_{-}}{E_{+} + E_{-} - i\hbar \Omega}
[n_{F}(\mu- E_+) - n_F(\mu + E_-)] \right\},
\end{split}
\end{equation}
where $n_F$ is the Fermi function, $A_{\pm} \equiv 1 \pm
\frac{\hbar^2v_F^2 \mathbf{k}_+ \mathbf{k}_- + \Delta^2}{E_+ E_-}
$, $\mathbf{k}_{\pm} = \mathbf{k} \pm \mathbf{K}/2$, and $E_{\pm}
= \sqrt{\hbar^2v_F^2 \mathbf{k}_{\pm}^2 + \Delta^2}$. Here $v_F$
is the Fermi velocity, $\mu$ is the chemical potential and
$\Delta$ is the excitonic gap that may open due to the interaction
between quasiparticles \cite{Khveshchenko-Gorbar}.
Eqs.~(\ref{LF.res}), (\ref{LF.res+}) and (\ref{LF.res-}) are
obtained using an effective long-wavelength description of the
semimetallic energy band structure of a single graphene layer with
the dispersion law $E(\mathbf{k}) = \pm \sqrt{\hbar^2v_F^2 k^2 +
\Delta^2} + \mu$ which allows a simple description in terms of the
Dirac equation in 2D (see
Refs.~\cite{Gonzalez:1999:PRB,Gonzalez:2001:PRB,Khveshchenko-Gorbar}).
In the absence of interactions, in undoped graphite, the
conduction and valence bands are touching each other in the two
inequivalent $K$ points. This corresponds to choosing $\mu =
\Delta =0$, i.e. in undoped graphite, at $T=0$ the negative band
$E= -\hbar v_F k$ is filled and the positive one $E= \hbar v_F k$
is empty. Notice that recent measurements of quantum magnetic
oscillations in graphite \cite{Luk'yanchuk:2004} confirm the
presence of holes with 2D Dirac-like spectrum $E = \pm \hbar v_F
k$, and this seems to be responsible for the strongly-correlated
electronic phenomena in this material. The effects of doping and
opening of the excitonic gap in graphene can be taken into account
by considering nonzero chemical potential $\mu$ and nonzero excitonic
gap $\Delta$, respectively.

Eqs.~(\ref{LF.res}), (\ref{LF.res+}) and (\ref{LF.res-}) underline the most
important difference between graphene and the usual two dimensional electron gas (see e.g.
Refs.~\cite{Tsvelik,Irene:2003:PRB}): while in the latter only the
$\chi_+$ term is present (intra-band excitations), in the graphene response also the
$\chi_-$ term contributes: this reflects the possibility of
particle-hole excitations with the particle and hole belonging
to different (positive and negative) branches of the spectrum ({\it inter-band} excitations).

Actually when $\mu =k_B T =0$  only the ``unusual'' $\chi_-$ term survives,  and one can
obtain an exact expression for $\chi(i\Omega_n \to \omega +i 0,K)$
\cite{Pisarski:1984:PRD}
which for $\Delta =0$ takes a rather simple form
\cite{Gonzalez:1999:PRB,Gonzalez:2001:PRB}
\begin{equation}\label{LF.exciton}
\begin{split}
&\chi(\omega,K;T=0) = \chi_-(\omega,K)
= - \frac{K^2}{4} \\
& \times \left[ \frac{ \theta(\hbar v_F K -
|\hbar\omega|)}{\sqrt{\hbar^2v_F^2 K^2 - \hbar^2\omega^2}} +i
\frac{\mbox{sgn}(\omega) \theta(|\hbar\omega|-\hbar v_F
K)}{\sqrt{\hbar^2\omega^2 - \hbar^2v_F^2 K^2}}\right].
\end{split}
\end{equation}

For finite $T$ the response function acquires an {\it additional
contribution} from $\chi_+$. We find an analytic expression for
$\chi_+$ in the limit  of small $\omega$ and $K$ but {\it finite}
$T$ and $\mu$ \cite{note1}. In fact considering
Eq.~(\ref{LF.res+}), in such a limit one obtains $A_{+} \approx
2$, so that, setting $\Delta = 0$, we obtain
\begin{equation}\label{LF.mu&T}
\begin{split}
&\chi_+(\omega,K) \simeq -\frac{4}{\pi} \frac{k_BT}{\hbar^2v_F^2} \ln \left(2 \cosh
\frac{\mu}{2k_BT}\right) \\
& \times \left[ 1 - \frac{|\hbar\omega| \theta(|\hbar\omega| - \hbar v_F
K)}{\sqrt{\hbar^2\omega^2 - \hbar^2v_F^2 K^2}} +
i \frac{\hbar\omega \theta(\hbar v_F K -|\hbar\omega|)}{\sqrt{\hbar^2v_F^2 K^2 - \hbar^2
\omega^2}} \right].
\end{split}
\end{equation}
Eq.~(\ref{LF.mu&T}) is one of the results of the present paper.

For $\mu \gg k_B T$ Eq.~(\ref{LF.mu&T}) reduces to the Lindhard
function of the two dimensional electron gas (2DEG) \cite{Tsvelik,Irene:2003:PRB},
while for $\mu =0$ the prefactor before the square brackets is
equal to $-(4 \ln 2/\pi) k_B T/\hbar^2 v_F^2$. Thus, as
anticipated, for $\mu = k_B T =0$ the only process that
contributes to nonzero $\mbox{Im} \chi$ is due to
Eq.~(\ref{LF.res-}).

Comparing Eqs.~(\ref{LF.exciton}) and (\ref{LF.mu&T}),
we notice that the domains where the
function $\mbox{Im} \chi_{\pm}(\omega, K)$,
which `counts' the number of particle-hole excitations of
energy $\omega$ and momentum $K$, is nonzero are different:
$|\omega| > v_F K$ for Eq.~(\ref{LF.exciton})
and $|\omega| < v_F K$ for Eq.~(\ref{LF.mu&T}).
At the formal level this difference in domains where particle-hole
excitations are allowed comes from the fact that
Eq.~(\ref{LF.mu&T}) describes processes where two involved
quasiparticles are from the same branch of the spectrum
(see Eq.~(\ref{LF.res+})), while Eq.~(\ref{LF.exciton}),
as mentioned above, originates from processes where two
quasiparticles belong to the different branches of the spectrum
(see Eq.~(\ref{LF.res-})).
Using the Dirac equation analogies, one can say that for nonzero
electron mass $\Delta$, Eq.~(\ref{LF.mu&T}) would describe electron-electron
scattering, while Eq.~(\ref{LF.exciton}) is related to
the process of electron-positron pair creation.

As predicted in Refs.~\cite{Flensberg:1994:PRL}
and experimentally confirmed (see e.g. Ref.~\cite{Rojo:1999:JPCM})
for $T \gtrsim 0.2 T_F$
(where $T_F$ is the Fermi temperature of the electron gas),
in metallic electron gases,
 the coupled
acoustic and optic plasmon modes occurring in the layered system
do contribute to the transresistivity resulting in its substantial
enhancement.

\begin{figure}
\putfig{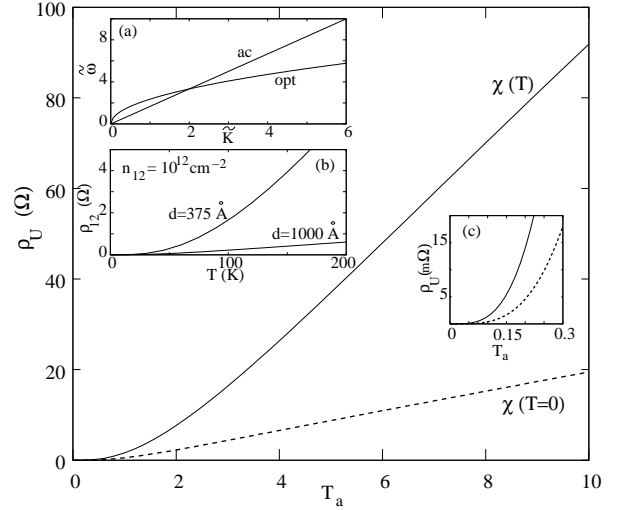}{8.0}
\caption{Universal transresistivity $\rho_U$ vs. dimensionless temperature $T_a$. The dashed line
corresponds to the approximated calculation using $\chi=\chi_-$. Inset (a): rescaled universal
plasmon dispersion relation $\tilde\omega$ vs. $\tilde K$. Inset (b): transresistivity $\rho_{12}$ vs.
temperature for two different value of the interlayer distance $d$ (as labelled).
Inset (c): as main panel, but for low values of $T_a$
In all calculations we have used $v_{F} =  9.7 \times
10^5m/s$ and $\varepsilon = 6$.
\cite{Gonzalez:2001:PRB}}
\label{rho}
\end{figure}

In the case of graphene we do expect such enhancement, since the
semimetallic band structure of graphite allows nonzero $\mbox{Im}
\chi(\omega,K)$ for $|\omega| > v_F K$ even for $T =0$. Due to
this peculiarity, and {\it at variance with usual 2DEG Coulomb
drag}, the transresistivity plasmon enhancement occurs already at
low temperatures, $T \gtrsim 0$ (see Fig.~\ref{rho}, inset (c)
where the transresistivity is plotted for low values of the
temperature rescaled by the characteristic Coulomb energy, see below).

Following the method described in Ref.~\cite{Rojo:1999:JPCM}, we have derived the plasmon dispersion
relations for small $K$ and $\omega$ ($|\omega| > v_F K$).
 Their expressions are given by
\ba
\hbar\omega^{ac}&=& 2\sqrt{\ln 2}\sqrt{{e^2\over\varepsilon d}k_BT}Kd ~~~~\mbox{and}\label{pl_ac}\\
\hbar\omega^{opt}&=& 2\sqrt{2}\sqrt{\ln
2}\sqrt{{e^2\over\varepsilon d}k_BT}\sqrt{Kd}\label{pl_opt} \ea
Notice that, apart from the factor $\sqrt{\ln 2}$,
Eqs.~(\ref{pl_ac}) and (\ref{pl_opt}) are formally similar to the
well known electron liquid expressions, except that the Fermi
energy $\mu$ ($\mu=0$ in the undoped system we are considering)
is substituted by the characteristic Coulomb energy  $E_C\equiv
e^2/(\varepsilon d)$. Therefore the system behavior will be in
this case characterized by the crossover between Coulomb and
thermal energy. This allows us to define the dimensionless
temperature $T_a=k_BT/E_C$ such that for $T_a>>1$ the system will
be in a non-degenerate regime, while for $T_a<<1$ the system will
be dominated by Coulomb interactions. This simple energy
scale-related analysis confirms that, even for small temperature
($T_a<<1$), plasmon enhancement will be important in graphene
systems (see Fig.~\ref{rho}, inset (c)).

If we introduce the dimensionless quantities $\tilde\omega=\hbar \omega/\sqrt{E_Ck_BT}$,
$\tilde K=Kd$,  we notice that Eqs.~(\ref{pl_ac}) and (\ref{pl_opt}) acquire the
{\it universal} form
\ba
\tilde\omega^{ac}&=& 2\sqrt{\ln 2}\tilde K ~~~~\mbox{and}\label{pl_ac_a}\\
\tilde\omega^{opt}&=& 2\sqrt{2}\sqrt{\ln 2}\sqrt{\tilde
K}\label{pl_opt_a}, \ea
which is plotted in
Fig.~\ref{rho}, inset (a). If in addition we define the ``fine-structure''
constant $\alpha^{-1}=(\hbar v_F/d)/E_C = \hbar v_F \epsilon
/e^2$, we find an important {\it scaling property} of the
transresistivity in respect to the interlayer distance $d$:
Eq.~(\ref{transresistivity.def}) in fact acquires the form \be
\rho_{21}(T) = {1\over n_1 n_2 d^4}\rho_U
(T_a,\alpha),\label{rho_12} \ee where \ba  \rho_U
(T_a,\alpha)&\equiv& -\frac{\hbar}{e^2}{1\over
2}{\alpha^2\over\sqrt{T_a}} \int_0^\infty d\tilde K\tilde
K\exp(-2\tilde K) \times\nonumber\\& &  \int_0^\infty d\tilde
\omega {\mbox{Im }\tilde\chi_1(\tilde \omega,\tilde K;\alpha)
\mbox{Im}\tilde\chi_2(\tilde \omega,\tilde K;\alpha)\over
|\epsilon(\tilde \omega,\tilde K;\alpha)|^2 \sinh^2(\tilde
\omega/(2\sqrt{T_a}))}\nonumber\\
 & \equiv & -\frac{\hbar}{e^2}\int_0^\infty d\tilde K\int_0^\infty
d\tilde \omega I(\tilde \omega,\tilde K;\alpha),\label{rho_U}
 \ea
is a universal function
of $T_a$ and $\alpha$, {\it independent} of the interlayer
distance $d$.

In Fig.~\ref{rho} we plot $\rho_U$ as a function of $T_a$. The
solid line corresponds to use in Eq.~(\ref{rho_U}) the full
$T$-dependent response function $\chi(T)$ (sum of
Eqs.~(\ref{LF.exciton})
 and (\ref{LF.mu&T})); the dashed line corresponds instead
to the approximation Eq.~(\ref{LF.exciton}).
The effect of including $\chi_+$, i.e. including temperature dependent effects, is quite significant.
At variance with the usual Coulomb drag,  the strong transresistivity enhancement
 is due in this case both  to the enlargement of the
 single-particle phase-space -- which includes now
{\it intra-band} excitation --  and to
 collective optical and acoustic plasmons. All these excitations are forbidden at T=0 and
not included in $\chi_-$.

\begin{figure}
\putfig{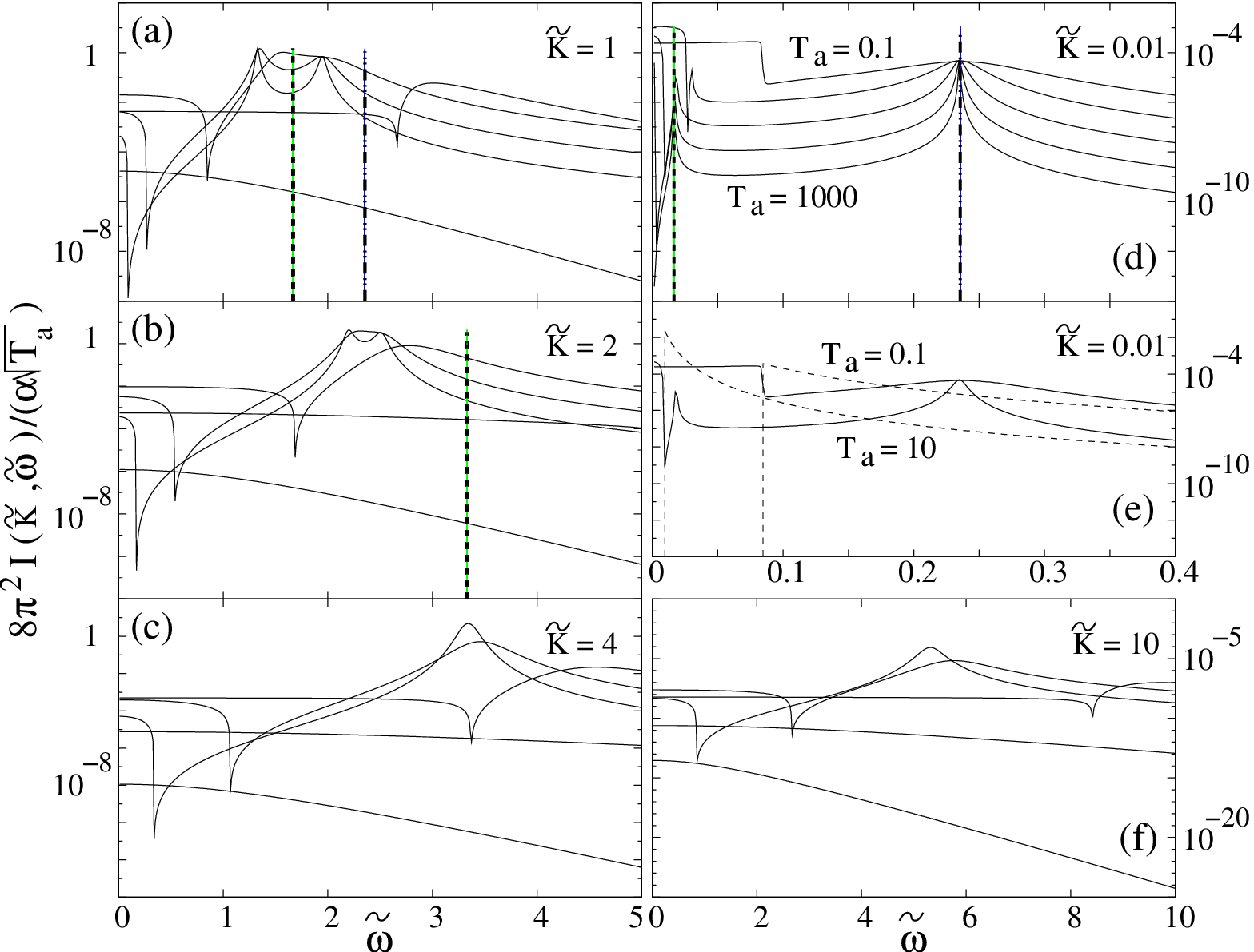}{8.5}
\caption{Transresistivity integrand $8\pi^2I(\tilde\omega,\tilde K)/(\alpha\sqrt{T_a})$ vs $\tilde\omega$ for
different values of $\tilde K$ (as labelled).
Panels (a), (b), (c), (d) and (f): Five values of $T_a$ (separated by a factor 10)
have been considered, $T_a=0.1\to1000$, each corresponding to a different solid line (see labelling in panel(d)).
Dashed (dashed-double-dot) line corresponds to analytical position of the acoustic (optical) plasmon.
Panel(e): Only two values of $T_a$ are considered (as labelled);
dashed lines correspond to calculations by using the approximation $\chi=\chi_-$.  }
\label{plasm}
\end{figure}

The inset (b) of Fig.~\ref{rho} presents the calculation of
$\rho_{21}(T)$ from Eq.~(\ref{transresistivity.def}) in respect to
temperature, for two different interlayer distance ($d=375
\mbox{\AA}$ and $d=1000 \mbox{\AA}$) and a typical 2DEG density.
This clearly shows that the dependence on $d$ of the
transresistivity is very strong. In this respect it is even more
remarkable that both curves of inset (b) collapse onto the
solid-line curve of the main panel when the reduced units are
considered. It would be very interesting to  check experimentally
such scaling behavior, which decouples the effect of layers'
separation from the interlayer Coulomb interaction.
Eq.~(\ref{rho_12}) in fact suggests that the main effect of
separating two interacting graphene layers by a distance $d$ is to
renormalize the layer electron densities according to such a
distance. We emphasize that, when considering usual Coulomb drag
experiments  between quantum wells, such effect would be hidden by
unavoidable experimental fluctuations  of different parameters,
such as finite quantum well thicknesses or doping. To check our
predictions, we suggest a Coulomb drag experiment between two {\it
large radius} carbon nanotubes, coaxial \cite{Lunde:2004:SST}
 or with parallel axes.
In this way  the aforementioned
fluctuations would be automatically avoided allowing the
experimental observation of such a scaling property.
In the remaining part of the paper, we concentrate on the plasmon
behavior and on the effect of this scaling property on plasmons.
In Fig.~\ref{plasm} we plot the rescaled integrand $8\pi^2I(\tilde
\omega,\tilde K;\alpha)/(\alpha\sqrt{T_a})$ (see
Eq.~(\ref{rho_U})) as a function of $\tilde \omega$ for different
$\tilde K$ and $T_a$. Let us first focus on Fig.~\ref{plasm}(d).
 It shows  the low $\tilde K$ regime, in which the
approximate expressions Eqs.~(\ref{pl_ac_a}) and (\ref{pl_opt_a})
describing the plasmons hold. The dimensionless temperature $T_a$
is varied by a factor 10 between $T_a=0.1$ and $T_a=1000$, as
labelled \cite{note2}. The vertical dashed-double-dot line
corresponds to the position of the optical plasmon (from
Eq.~(\ref{pl_opt_a})), and the dashed line to the acoustic plasmon
one (from Eq.~(\ref{pl_ac_a}). Notice that plasmons are present in
the domain  $|\omega|>\hbar v_F$ only, so if the value of $\tilde
\omega$ corresponding to the plasmon does not belong to such a
region, the corresponding plasmon disappears. This is the case for
the acoustic plasmon at $T_a=0.1$ and $T_a=1$. Fig.~\ref{plasm}(d)
clearly shows that the scaling law  Eqs.~(\ref{pl_ac_a}) and
(\ref{pl_opt_a}) are well satisfied, irrespectively of the value of
$T_a$. In addition, due to the rescaling
$8\pi^2/(\alpha\sqrt{T_a})$,
 the strength of the
optical plasmon becomes independent of $T_a$.

By inspecting Fig.~\ref{plasm}(a), (b) and (c), we see that the behavior predicted by
Eqs.~(\ref{pl_ac_a}) and (\ref{pl_opt_a}) is qualitatively respected even for $\tilde K$
of the order of unity, especially as long as the optical plasmon is considered. In particular
these equations predict an overlap of the two plasmons at $\tilde K=2$ (dashed line, panel(b)).
Indeed the two plasmons join
 for a reasonably close value of $\tilde K$ (see panel(c), $\tilde K=4$ ).
 For increasing $\tilde K$ (large
$\tilde K$ regime) though, a single plasmon-structure is found, i.e. the plasmon behavior
differs even qualitatively from Eqs.~(\ref{pl_ac_a}) and (\ref{pl_opt_a}) (see panel (f),
$\tilde K=10$).

Fig.~\ref{plasm}(e) compares $8\pi^2I(\tilde \omega,\tilde K;\alpha)/(\alpha\sqrt{T_a})$
calculated using $\chi=\chi_-+\chi_+$ (solid line),
to the approximation  obtained using Eq.~(\ref{LF.exciton}) (dashed lines),
 for two values of $T_a$.
As can be clearly seen, the integrand is influenced not only by
the plasmon structure, which  enhances it for $|\omega|>\hbar
v_F$, but also by the single particle intraband excitation continuum: for $|\omega|<\hbar
v_F$, the integrand  would be otherwise vanishing (see dashed lines).

In this Letter we have presented the first, to the best of our
knowledge, calculation of Coulomb drag effects between graphene
layers, which include temperature-dependent effects. We have also
 predicted a universal behavior of the Coulomb
transresistivity,  which is interlayer-distance independent and
suggested that
this behavior may be experimentally observable.
Most of the calculations presented  were made for the simplest
gapless form of the quasiparticle spectrum in graphene, as
expected from tight-binding calculations in the absence of
interactions. These results, can be easily generalized to take
into account the possible opening of a dielectric gap $\Delta$ in
the quasiparticle spectrum
\cite{Kopelevich:2002,Khveshchenko-Gorbar}. The interplay of this
gap with the temperature and chemical potential may lead to even
richer physics and applications.

\section{Acknowledgments}
We thank V.P.~Gusynin, V.M.~Loktev and J.~Dobson for helpful
discussions. S.G.Sh. is also grateful to the members of ISI for
the friendly support  and hospitality.


\begin{thebibliography}{99}
\bibitem{note1}
By inspecting Eq.~(\ref{transresistivity.def}), we then expect
that our results for the drag transresistivity will be more
reliable for relatively small temperatures ($\hbar\omega
\stackrel{<}{~}k_BT$) and finite interlayer distance $d$
($K\stackrel{<}{~}d^{-1}$).
\bibitem{note2} We underline that, depending on the different $d$ considered, the same
dimensionless temperature may correspond to low or high
temperatures $T$ -- for example for $d=1000 \mbox{\AA}$ and $d=375
\mbox{\AA}$, $T_a=1$ corresponds to $T=28$K and $T=74$K
respectively.
\end{thebibliography}
\end{document}